\documentclass{article} 
\usepackage{gmas_iclr2022_conference,times}

\usepackage{amsmath,amsfonts,bm}

\def\eqref#1{equation~\ref{#1}}

\def\1{\bm{1}}

\DeclareMathAlphabet{\mathsfit}{\encodingdefault}{\sfdefault}{m}{sl}
\SetMathAlphabet{\mathsfit}{bold}{\encodingdefault}{\sfdefault}{bx}{n}

\usepackage{hyperref}
\usepackage{url}
\usepackage{enumitem}
\usepackage{caption}
\usepackage{subcaption}
\usepackage{graphicx}
\usepackage{booktabs}

\title{Dynamic Noises of Multi-Agent Environments Can Improve Generalization: Agent-based Models meets Reinforcement Learning}

\author{Mohamed Akrout, Amal Feriani \& Bob McLeod \\
Department of Electrical and Computer Engineering\\
University of Manitoba\\
Winnipeg, MB, Canada \\
\texttt{\{akroutm, feriania\}@myumanitoba.ca}, \texttt{\{robert.mcleod\}@umanitoba.ca} \\
}

\iclrfinalcopy 
\begin{document}

\maketitle

\begin{abstract}
We study the benefits of reinforcement learning (RL) environments based on agent-based models (ABM). While ABMs are known to offer microfoundational simulations at the cost of computational complexity, we empirically show in this work that their non-deterministic dynamics can improve the generalization of RL agents. To this end, we examine the control of an epidemic SIR environments based on either differential equations or ABMs. Numerical simulations demonstrate that the intrinsic noise in the ABM-based dynamics of the SIR model not only improve the average reward but also allow the RL agent to generalize on a wider ranges of epidemic parameters.
\end{abstract}
\section{Introduction}
\subsection{Background and related work}

Most dominant learning algorithms rely on single-agent and optimization-based paradigms. However, many intelligent systems in various domains such as economy, biology, gaming, ecology and sociology are multi-agent and/or decentralized in nature (\cite{ponomarev2017multi}). This calls for the need to $i)$ design systems accounting for the intrinsic interactions between its multiple components, and $ii)$ investigate the impact of this consideration in the overall performance and the generalization of the existing algorithms. 

Generalization has become one of the major challenges of real-world Reinforcement Learning (RL). Despite the tremendous success of RL in a myriad of applications, the learned policies have difficulty generalizing to new test environments, even when these environments are slightly different from the training ones (\cite{Farebrother2018, Zhang2018, CobbeKHKS19, SongJTDN20}). Indeed, RL agents are commonly trained and evaluated on the same environment, thereby leading to overfitting and biased evaluation (\cite{WhitesonTTS11}). This practice is fundamentally different from supervised learning settings where the optimization and the evaluation are conducted on two separate sets (i.e., train and test sets). Empirical risk minimization guarantees a good generalization for supervised learning in absence of data distributional shift. This motivated the construction of similar settings in RL to improve generalization: The agent learns using a training set of contexts or tasks and is evaluated on new unseen test contexts (\cite{CobbeKHKS19}).

The existing RL suites for generalization such as OpenAI Procgen (\cite{CobbeHHS20}) and 2D CoinRun (\cite{CobbeKHKS19}) are solely devoted to quantify generalization in vision-based RL tasks. To do so, a set of game levels are procedurally generated by varying visual styles and level layouts without changing the environment dynamics. However, little attention has been given for non-vision RL tasks where the agent is supposed to generalize across different dynamics (\cite{packer2018assessing, PengAZA18}). This is because artificially perturbing the dynamics of complex non-linear dynamical systems (e.g., fluid dynamics, electromagnetism, etc.) requires a careful design of the dynamics randomization procedure to avoid unrealistic scenarios (e.g. deterministic chaos, strange attractors). Such problems are challenging for traditional and deep control methods to handle (\cite{bucci2019control}).

\subsection{Motivation}

\begin{figure}[t]
     \begin{subfigure}[b]{0.4\textwidth}
         \includegraphics[scale=0.75]{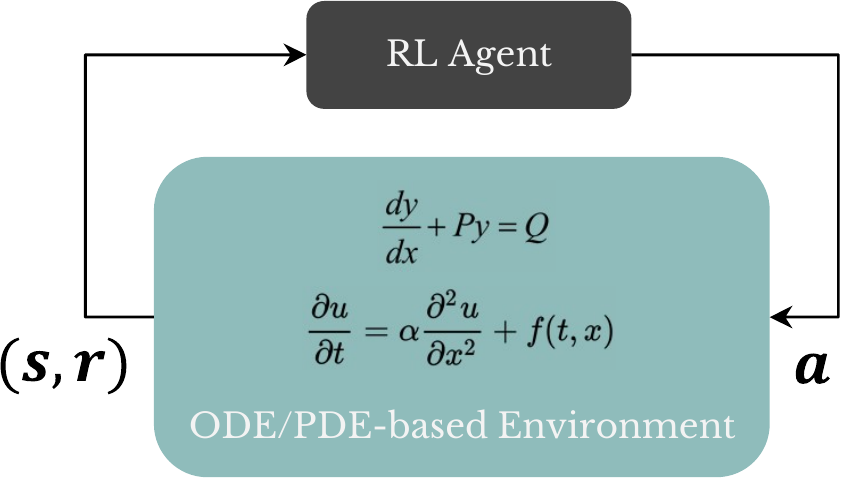}
         \caption{}
         \label{fig:ode-env}
     \end{subfigure}
     \hspace{1.5cm}
     \begin{subfigure}[b]{0.4\textwidth}
         \includegraphics[scale=0.75]{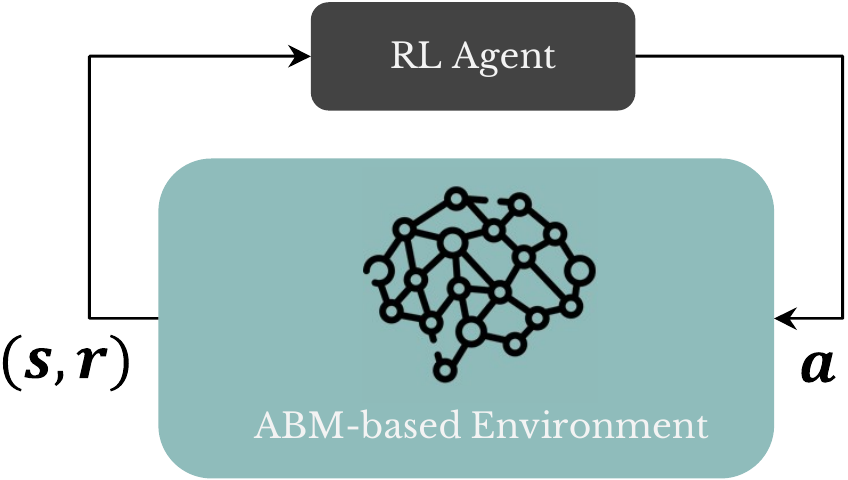}
         \caption{}
         \label{fig:abm-env}
     \end{subfigure}
     \caption{The RL setting using RL environments whose dynamics are simulated using (a) ODEs and PDEs, and (b) ABMs.}
     \label{fig:ode-abm-envs}
\end{figure}

As depicted in Fig.~\ref{fig:ode-abm-envs}, we advocate in this work the use of agent-based models (ABM) to design a class of environments for dynamics generalization without the injection of artificial noise. The motivation behind this work stems from the fact that several real-world applications in social and network sciences rely on complex interactions between the components of the overall simulated system. The dynamics of the latter are out of reach for mathematical methods (\cite{epstein1996growing,colman1998complexity}), i.e., they cannot be explicitly governed by ordinary differential equations (ODE) or partial differential equations (PDE). This is in contrast to many real-word applications such as robotics (\cite{kober2013reinforcement}), ventilating (\cite{farahmand2016learning}), and fluid dynamics \cite{pirmorad2021deep} whose complex dynamical systems can be captured using ODEs and PDEs. Moreover, it has been shown that RL agents are prone to exploiting idiosyncrasies of specific implementations of RL environments from which they learn infeasible behaviors in the real world (\cite{heess2017emergence}). This fact justifies the need for more realistic simulators for RL environments based on ABMs despite their computational complexity.

\noindent ABMs come into play as a computational modelling approach for several dynamical systems based on the microfoundational simulation of heterogeneous interactions across predefined agents with heterogeneous preferences (\cite{bianchi2015agent}). ABMs capture emergent self-organizing phenomena induced by the interactions of the individuals, leading to a system behavior whose properties are decoupled of the agents' properties. This aspect commonly summarized by ``the whole is more than the sum of its parts'' is a modelling bottleneck because its explicit mathematical description is unknown and/or not easy to obtain.

Furthermore, several ODE/PDE-based RL environments tend to smooth out fluctuations by modelling the average behavior across all the agents. However, under certain conditions, fluctuations can be amplified when the system is linearly stable but unstable to larger perturbations. Therefore, simulating a population of agents using dynamic parameters computed from averaging over the behavior of all agents is an oversimplification (\cite{bonabeau2002agent}). Indeed, the behavior of each individual is nonlinear and highly dependent on the dynamics of its neighbors. For example, some behavioral scenarios can follow a hard-thresholding decision making strategies. The latter induce a discontinuity in the behavior that cannot be captured by differential equations.

Last but not least, the dynamics of ABMs are non-deterministic due to the randomness in the behavior of each agent. Hence, they induce intrinsic noises that are not artificially injected to examine the effect of the dynamic randomization on the generalization of the RL agent.

\subsection{Contributions}

The use of ABMs for the microfoundational simulations of the dynamics of RL environments is not new a idea (\cite{rand2006machine,sert2020segregation}). However, to our best knowledge, the effect of the intrinsic noise in the ABM dynamics on the generalization of RL agents has never been investigated. In this paper, we benchmark the generalization of an RL agent interacting with a standard epidemic SIR environment implemented with either ODEs~(\cite{kermack1927contribution}) or ABMs (\cite{perez2009agent}). We first compare the evolution of the SIR dynamics using ODEs and ABMs to seize the intrinsic noise in the dynamics of the ABM-based SIR environment. We then randomize the epidemic parameters of the SIR model during both training and inference. By doing so, we show that an RL agent obtains a significant improvement in the average reward and an ability to generalize across different ABM-based environments whose epidemic parameters correspond to distinct pandemic regimes. Our work present an empirical evidence about the relationship between ABM-based environments and RL generalization, thereby calling for the need to further investigate this relationship in future work.

\section{The ABM-based RL environment: the SIR model as a study case}
\subsection{The ODE-based SIR dynamics}\label{subsec:ode-based-sir}
\begin{figure}[h!]
    \centering
    \includegraphics[scale=0.6]{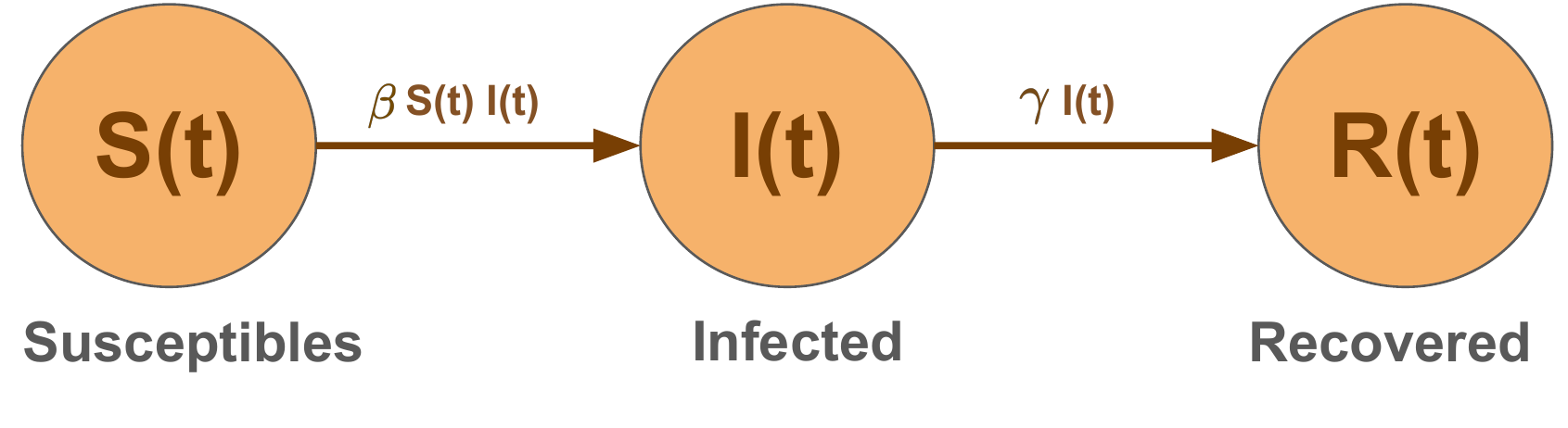}
    \caption{The finite state machine representation of the SIR model.}
    \label{fig:finite-state-machine}
\end{figure}

The spread of a disease can be modelled using the well-known SIR model \cite{kermack1927contribution}. The latter divides the (fixed) population of $N$ individuals into the following three varying group of individuals over the time $t$ (measured in days):
\begin{itemize}[leftmargin=*]
    \item $S(t)$ is the number of susceptible but not yet infected individuals,
    \item $I(t)$ is the number of infected individuals,
    \item $R(t)$ is the number of recovered individual from the disease, and hence are naturally immunized.
\end{itemize}
Using the finite state machine representation, these three groups are associated to three states as shown in Fig.~\ref{fig:finite-state-machine}
The transition between them is governed by two key parameters:
\begin{itemize}[leftmargin=*]
    \item $\beta$ represents the effective contact rate of the disease. At every time step $t$, an infected individual in the group $I(t)$ comes into contact with a sub-population of size $\beta\,N$ corresponding to the fraction $S(t)/N(t)$ of individuals that are susceptible to contract the disease.
    \item $\gamma$ is the mean recovery rate. This corresponds to a mean period of time, $\gamma^{-1}$, during which an infected individual among the group $I(t)$ can spread the virus to its contacts.
\end{itemize}

The dynamics of this system can also be formulated using the following set of differential equations~(\cite{kermack1927contribution}):

\begin{equation}
\begin{aligned}
    \frac{\text{d}S(t)}{\text{d}t} &= - \beta\,\frac{S(t)\,I(t)}{N},\\
    \frac{\text{d}I(t)}{\text{d}t} &= \beta\,\frac{S(t)\,I(t)}{N} - \gamma\, I(t),\\
    \frac{\text{d}R(t)}{\text{d}t} &= \gamma\,I(t).
\end{aligned}
\end{equation}

\subsection{The ABM-based SIR dynamics}\label{subsec:abm-based-sir}
We consider a complete undirected graph between $N$ homogeneous agents. Each agent can be in one of the three possible SIR states in Fig~\ref{fig:finite-state-machine}. At each time step, if the agent is infected, it transmits the disease with a probability $p_{\text{trans}}=\beta/N$ to each neighboring agent in the susceptible state. Moreover, each infected agent transits to the recovered state after $\gamma^{-1}$ time steps (in days).


The choice of the complete graph topology stems from the fact that it captures more accurately the idea of the aggregate population. In other words, the more interactions between agents are allowed, the closer the average dynamics of the ABM becomes as compared to the ODE-based one.




\subsection{The intrinsic noise in the ABM-based SIR dynamics}

\begin{figure}[h!]
\centering
\includegraphics[scale=.53]{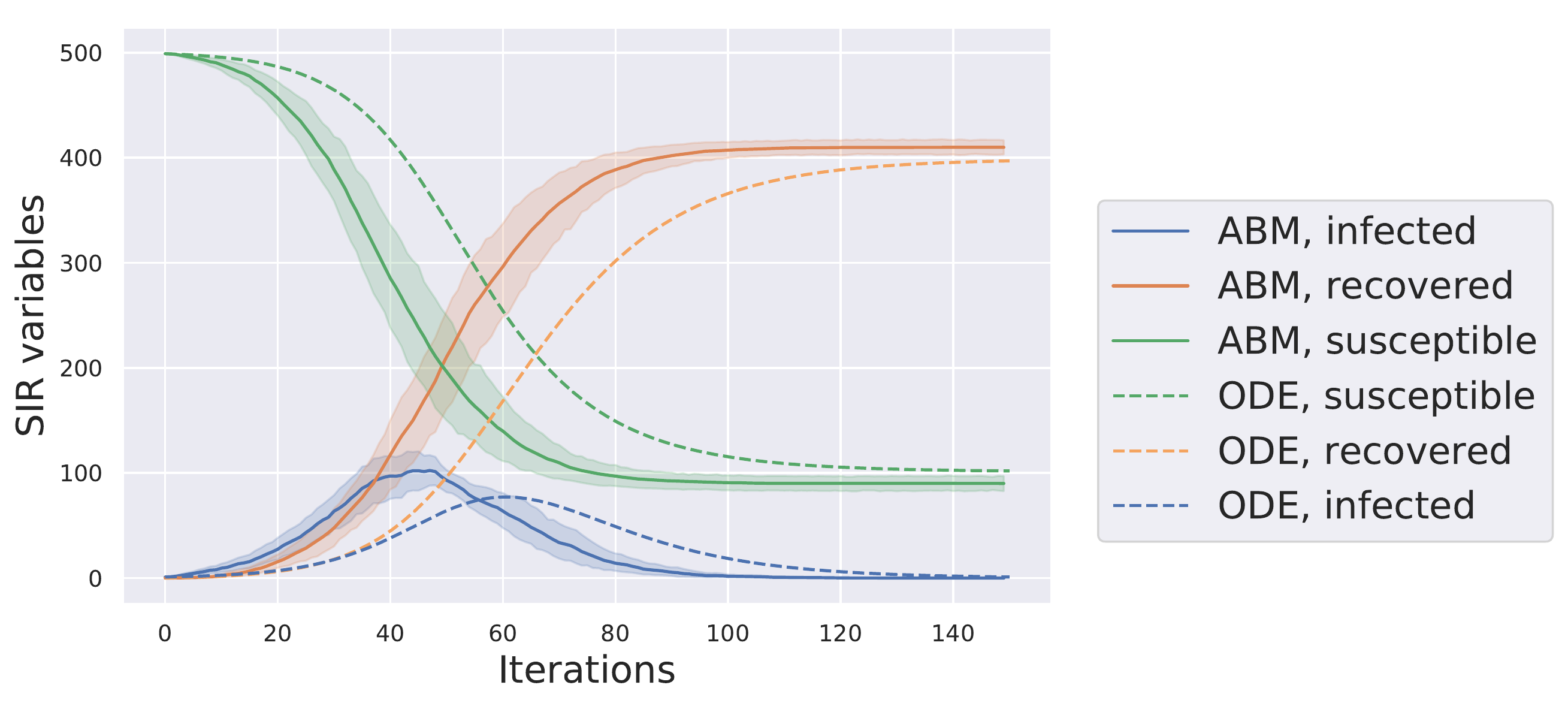}
\caption{The state evolution of the SIR model as a function of iterations for a population size $N=500$, $\beta=0.2$ and $\gamma=0.1$ when simulated with ODEs (in dashed line) and 100 Monte Carlo realizations of ABMs (in solid line).}
\label{fig:evolution-SIR-variables-ABM-vs-ODE}
\end{figure}

\noindent To verify our implementation of the SIR model with an ABM, we compare the ABM-based dynamics vs. the ODE-based one. Fig. \ref{fig:evolution-SIR-variables-ABM-vs-ODE} depicts the time evolution of the number of individuals belonging to the susceptible, infected, and recovered populations, namely, $S(t)$, $I(t)$ and $R(t)$, respectively. There, it is seen that the values produced by the ODE model (in solid lines) and the ABM (in dashed lines) converge to approximately the same values. This validates the asymptotic behavior of the ABM model against the ODE model. We also observe a discrepancy between the time evolution of the ABM-based and ODE-based values. This is due to the realistic discrete behavior of the agents as opposed to the averaged one from the ODE. The ``microscopic modelling'' of ABMs lead to significant deviations from predicted aggregate behavior based on the coarse-grained models (\cite{moein2021inefficiency}).\vspace{0.5cm}

The observed deviations from one single simulation to another can be seen as an intrinsic noise in the ABM-based dynamics shown in Fig.\ref{fig:evolution-SIR-variables-ABM-vs-ODE}. This non-deterministic aspect of ABM simulations are different from the artificial noise incorporation to deterministic ODE-based dynamics in order to study generalization in RL.

\section{Simulation results}

In this section, we empirically show how the intrinsic noise in ABM-based dynamics can be beneficial for RL generalization. To this end, we consider two pandemic propagation regimes: a low-spreading one with $\beta=0.2$ and a wide-spreading one with $\beta=0.8$.

We train an RL agent using proximal policy optimization (PPO) (\cite{schulman2017proximal})  to choose an action from the set $\mathcal{A}=\{\text{lockdown},\,\text{social distancing},\,\text{open}\}$ to control three SIR models:
\begin{itemize}[leftmargin=*]
    \item \textbf{ODE} : an ODE-based SIR model as described in subsection \ref{subsec:ode-based-sir};
    \item \textbf{Randomized ODE}: an ODE-based SIR model with dynamics randomisation at each time-step of each episode, where the SIR state ($S(t)$, $I(t)$, $R(t)$) is perturbed with a noise sampled from a uniform probability mass distribution between $1$ and $10$ individuals.
    \item \textbf{ABM}: an ABM-based SIR model as described in subsection \ref{subsec:abm-based-sir}.
\end{itemize}

We train and evaluate the RL agent interacting with each one of the aforementioned environments with the default parameters of the PPO implementation of the stable baselines (\cite{stable-baselines3}). We evaluate the RL agent on 5000 episodes. The number of evaluation episodes was chosen such as the evaluation time of ABMs does not exceed $12$ hours.

To investigate the learned policies of the RL agents, we present in Tables~\ref{tab:actions-per-method-high-pandemic} and \ref{tab:actions-per-method-low-pandemic} the distribution of the actions for wide- and low- spreading pandemics, respectively. We also report the average reward during the evaluation in Table~\ref{tab:average-reward}.

\begin{table}[h!]
\centering
\caption{The percentage of the taken actions for RL environments simulated based on ODEs, randamized ODEs and ABM for wide-spreading pandemic $\beta=0.8$.}
\begin{tabular}[t]{lccc}
\toprule
\textbf{}& \textbf{Lockdown}& \textbf{Social distancing}& \textbf{Open}\\
\midrule
ODE &  4.2\% & 0.9\%& 94.9\%\\ \midrule
Randomized ODE & 99.3\% & 0.2\%  & 0.5\% \\  \midrule
ABM & 0.7\% & 20.1\% &  79.1\%\\
\bottomrule
\end{tabular}
\label{tab:actions-per-method-high-pandemic}
\end{table}

\begin{table}[h!]
\centering
\caption{The percentage of the taken actions for RL environments simulated based on ODEs, randamized ODEs and ABM for low-spreading pandemic $\beta=0.2$.}
\begin{tabular}[t]{lccc}
\toprule
\textbf{}& \textbf{Lockdown}& \textbf{Social distancing}& \textbf{Open}\\
\midrule
ODE &  0.8\% & 42.1\%& 57.1\%\\ \midrule
Randomized ODE & 99.2\% & 0.4\%  & 0.4\% \\  \midrule
ABM & 0.4\% & 0.3\% &  99.3\%\\
\bottomrule
\end{tabular}
\label{tab:actions-per-method-low-pandemic}
\end{table}

\begin{table}[h!]
\centering
\caption{The average reward obtained with different RL environments for both low-spreading ($\beta=0.2$) and wide-speading ($\beta=0.8$) pandemics.}
\begin{tabular}[t]{cccc}
\toprule
\textbf{}& \textbf{ODE}& \textbf{Randomized ODE}& \textbf{ABM}\\
\midrule
$\beta=0.2$ &  0.77 $\pm$0.12 & -0.16 $\pm$ 0.15 & \textbf{0.84 $\pm$ 0.14}\\ \midrule
$\beta=0.8$ & 0.78 $\pm$0.8 & -0.15 $\pm$ 0.13 & \textbf{0.85 $\pm$ 0.11} \\
\bottomrule
\end{tabular}
\label{tab:average-reward}
\end{table}

We observe that the uniform randomisation of the ODE dynamics has drastically shifted the action distribution of the RL agent and led to a poor average reward in both pandemic regimes. This confirms that the blind randomization of the dynamics cannot be beneficial without domain knowledge, unlike visual-based randomization (\cite{lee2019network}).

Moreover, we observe that the RL agent trained on an ABM-based environment achieves a high average reward in the evaluation. Its policy has shifted from being ignorant of the pandemic dynamics (i.e., almost picks the action ``open'') in the low-spreading pandemic to a policy that consider social distancing with a frequency of $20\%$ in the wide-spreading regime. This suggests that the noise introduced by the non-deterministic dynamics of ABMs from one episode to another has improved the overall averaged reward. Further experiments should be conducted over a higher number of Monte Carlo simulations to validate this fact.

\begin{figure}[h!]
\centering
\includegraphics[scale=.53]{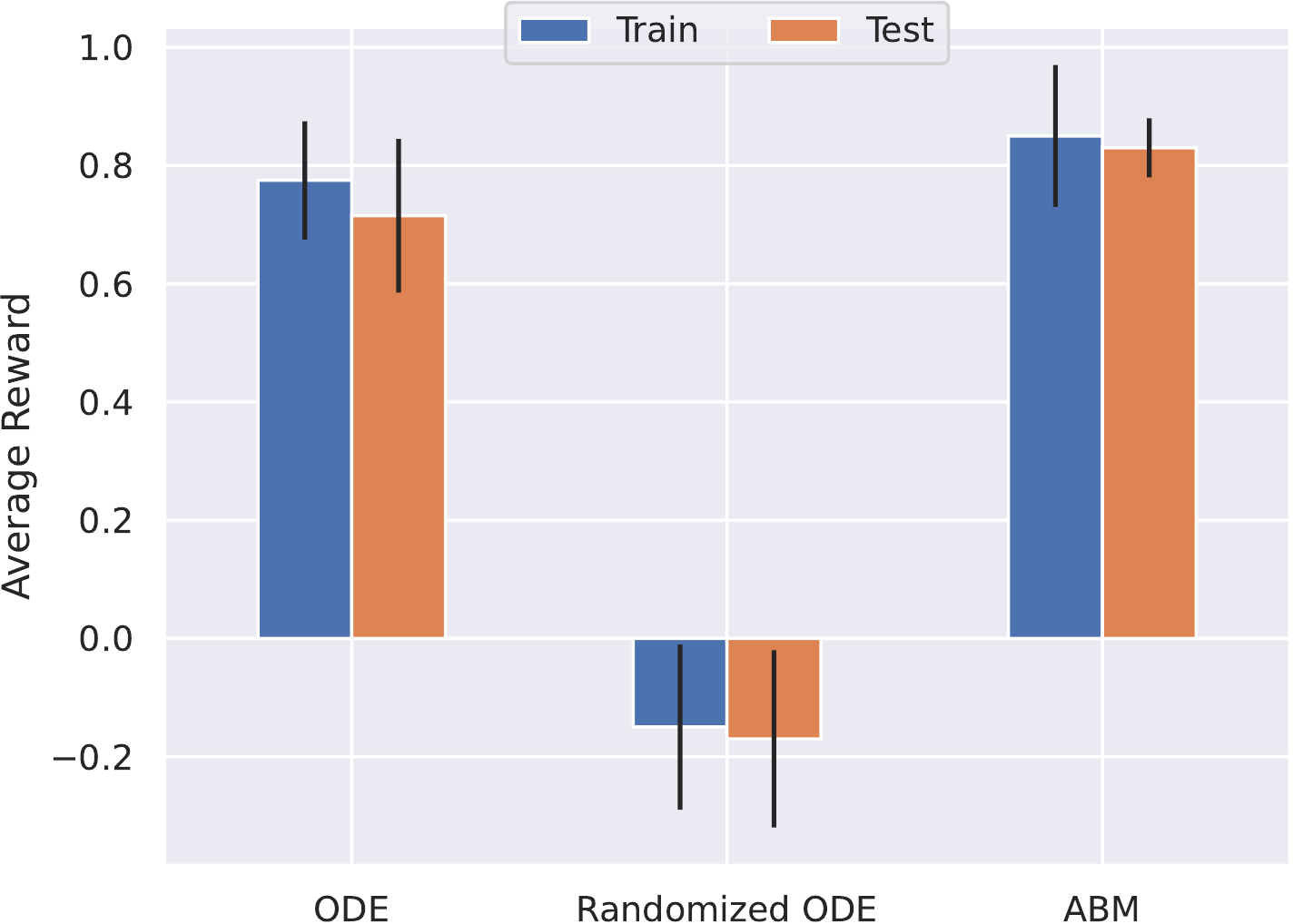}
\caption{The mean and standard deviation of the average reward over pandemic regimes with $\beta$ sampled uniformly in $[0.2,\,0.8]$, while $\gamma = 0.1$.}
\label{fig:generalization}
\end{figure}

Finally, we evaluate the generalization of the trained RL agents on test pandemic regimes with different $\beta$ values. To this end, we randomly sample the values of the parameter $\beta$ between $0.2$ and $0.8$ and run the learned policies on the train regimes (i.e., low and wide spreading pandemic regimes). Figure \ref{fig:generalization} illustrates the obtained results. It shows that the agent trained with ABM-based environments is able to generalize to different pandemic parameters and maintain similar performance (i.e., $2.3\%$ decrease) as in the train regimes. However, using ODE-based environments, we observe that there is a decrease in the performance (approximately $8\%$). This confirms that ABMs can offer a better design of RL environments not only for the sake of realistic modelling but also for RL generalization because of their non-deterministic dynamics. Moreover, our results highlight that artificially injecting noise in the agent state did not improve generalization (i.e., negative average reward) on the contrary to the common practice in generalization and sim-to-real works, e.g., (\cite{PengAZA18}).

\section{Conclusion}
In this work, we empirically demonstrated that ABM-based modeling improves the performance and generalization of RL agents due to the intrinsic noise of their dynamics. Our experiments showed that dynamics randomization through random perturbation leads to deterioration of the agent's performance, thereby calling for a more careful design of simulated environments for better generalization. We therefore advocated ABM-based modeling as an approach to tackle dynamics randomization without artificial noise injection during training. 

\bibliography{gmas_iclr2022_conference}
\bibliographystyle{gmas_iclr2022_conference}


\end{document}